\documentclass[aps,prd,12pt,preprint,tightenlines,superscriptaddress,
amsfonts,amssymb,amsmath,byrevtex,showpacs,nofootinbib]{revtex4}
\usepackage{graphics}
\usepackage{epsfig,subfigure}
\newcounter{mnotecount}[section]
\renewcommand{\themnotecount}{\thesection.\arabic{mnotecount}}
\newcommand{\mnote}[1]
{\protect{\stepcounter{mnotecount}}$^{\mbox{\footnotesize  $
      \bullet$\themnotecount}}$ \marginpar{\raggedright\tiny
    $\!\!\!\!\!\!\,\bullet$\themnotecount: #1} }

\begin{document}
\newcommand{\dR}{\mathbb R}
\newcommand{\dC}{\mathbb C}
\newcommand{\dZ}{\mathbb Z}
\newcommand{\dN}{\mathbb N}
\newcommand{\id}{\mathbb I}

\title{Turning Big Bang into Big Bounce: I. Classical Dynamics}

\author{Piotr Dzier\.{z}ak$^\S$, Przemys{\l}aw Ma{\l}kiewicz$^\dag$ and
W{\l}odzimierz Piechocki$^\ddag$
\\ Theoretical Physics Department, Institute for Nuclear Studies
\\ Ho\.{z}a 69, 00-681 Warsaw, Poland;
\\ $^\S$pdzi@fuw.edu.pl, $^\dag$pmalk@fuw.edu.pl, $^\ddag$piech@fuw.edu.pl}

\date{\today}

\begin{abstract}
The  big bounce (BB)  transition within a flat
Friedmann-Robertson-Walker  model is analyzed in the setting of
loop geometry  underlying the loop cosmology. We solve the
constraint of the theory at the classical level to identify
physical phase space and find the Lie algebra of the Dirac
observables. We express energy density of matter and geometrical
functions in terms of the observables. It is the modification of
classical theory by the loop geometry that is responsible for BB.
The classical energy scale specific to BB depends on a parameter
that should be fixed either by cosmological data or determined
theoretically at quantum level, otherwise the energy scale stays
unknown.
\end{abstract}
 \maketitle

\section{Introduction}
It is commonly believed that the cosmological {\it singularity}
problem \cite{MTW,JPAK,JMMS} may be resolved  in a theory which
unifies gravity and quantum physics. Recent analyses done within
the {\it loop} quantum cosmology (LQC) concerning homogeneous
isotropic universes of the Friedmann-Robertson-Walker (FRW) type,
strongly suggest that the evolution of these universes does not
suffer from the classical singularity. Strong  {\it quantum}
effects at the Planck scale cause that classical big bang is
replaced by quantum big bounce (BB)
\cite{Bojowald:2001xe,Date:2004fj,Ashtekar:2006rx,Ashtekar:2006wn,Ashtekar:2007em}.

The resolution of the cosmic singularity problem offered by LQC
requires the existence of a fundamental {\it length}, which
effectively implies the discreteness of quantum geometry. However,
the {\it size} of this length has not been determined satisfactory
yet, i.e. derived within LQC. Presently, it is an {\it ad-hoc}
assumption of standard LQC \cite{Dzierzak:2008dy,Bojowald:2008ik}.

Our paper is an extended version of the {\it classical} part of
\cite{Malkiewicz:2009zd}. Its goal is the demonstration that the
resolution of the initial cosmological singularity is  due to the
{\it modification} of the  classical theory by the  loop geometry.
Quantum effects seem to be of a secondary importance, but should
be examined since the energy scale specific to BB has not been
identified yet \cite{Dzierzak:2008dy,Malkiewicz:2009zd}. The big
bounce may occur deeply inside the Planck scale where it is
commonly expected that quantum effects cannot be ignored.

The difference between {\it standard} LQC
\cite{Ashtekar:2003hd,Bojowald:2006da} and our {\it nonstandard}
LQC is the following: (i) we determine an {\it algebra} of
observables on the kinematical phase space; (ii) we {\it solve}
the Hamiltonian constraints at the classical level to identify
{\it physical} phase space (i.e. the space of Dirac's
observables); (iii) we express functions on the physical phase
space, like the matter density and geometrical operators (length,
area, volume), in terms of Dirac's observables and an {\it
evolution} parameter; and (iv) by {\it quantization} we mean:
finding a self-adjoint representation of the algebra of the Dirac
observables and solution to the eigenvalue problem for operators
corresponding to functions specified in (iii). Roughly speaking,
in standard LQC one does not identify an {\it algebra} of physical
observables and one imposes the Hamiltonian constraints {\it only}
at the quantum level.

The present paper concerns the {\it classical} level so we are
mainly concern with items (i)-(iii). Our next paper \cite{all}
will be an extended version of the {\it quantum} part of
\cite{Malkiewicz:2009zd}, i.e. it will be devoted to the
realization of item (iv).

For simplicity of exposition we restrict ourselves to the  flat
FRW model with massless scalar field. This model of the universe
includes the initial cosmological singularity and has been
intensively studied recently within LQC.

In order to have our paper self-contained,  we recall in Sec. II
the form of classical Hamiltonian in terms of holonomy-flux
variables.  In Sec. III we solve the constraint and analyze the
relative dynamics in the {\it physical} phase space. Section IV is
devoted to the algebra of {\it observables}. We consider the
energy density of the scalar field and geometrical operators in
Sec. V. Higher order holonomy corrections are shortly discussed in
the appendix. We conclude in the last section.

\section{Hamiltonian}

The gravitational part of the classical Hamiltonian, $H_g$, in
general relativity is a linear combination of the first-class
constraints, and reads
\cite{TT,CR,Ashtekar:2004eh,Ashtekar:2003hd,Bojowald:2006da}
\begin{equation}\label{ham1}
    H_g:= \int_\Sigma d^3 x (N^i C_i + N^a C_a + N C),
\end{equation}
where $\Sigma$ is the spacelike part of spacetime $\dR \times
\Sigma$, $~(N^i, N^a, N)$ denote Lagrange multipliers, $(C_i, C_a,
C)$ are the Gauss, diffeomorphism and scalar constraint functions.
In our notation  $(a,b = 1,2,3)$ are spatial and $(i,j,k = 1,2,3)$
internal $SU(2)$ indices. The constraints must satisfy a specific
algebra.

Having fixed local gauge and diffeomorphism freedom we can rewrite
the gravitational part of the classical Hamiltonian (for the  flat
FRW model with massless scalar field) in the form (see, e.g.
\cite{Ashtekar:2006wn})
\begin{equation}\label{hamG}
H_g = - \gamma^{-2} \int_{\mathcal V} d^3 x ~N
e^{-1}\varepsilon_{ijk}
 E^{aj}E^{bk} F^i_{ab}\, ,
\end{equation}
where  $\gamma$ is the Barbero-Immirzi parameter, $\mathcal
V\subset \Sigma$ is an elementary cell, $\Sigma$ is spacelike
hyper-surface,  $N$ denotes the lapse function,
$\varepsilon_{ijk}$ is the alternating tensor, $E^a_i $ is a
densitized  vector field, $e:=\sqrt{|\det E|}$, and where
$F^i_{ab}$ is the curvature of an $SU(2)$ connection $A^i_a$.

The resolution of the singularity, obtained within LQC, is based
on rewriting  the curvature $F^k_{ab}$ in terms of holonomies
around loops.   The curvature $F^k_{ab}$   may be determined
\cite{Ashtekar:2006wn} by making use of the formula (see the
appendix)
\begin{equation}\label{cur}
F^k_{ab}= -2~\lim_{Ar\,\Box_{ij}\,\rightarrow \,0}
Tr\;\Big(\frac{h^{(\mu)}_{\Box_{ij}}-1}{\mu^2
V_o^{2/3}}\Big)\;{\tau^k}\; ^o\omega^i_a  \; ^o\omega^j_a ,
\end{equation}
where
\begin{equation}\label{box}
h^{(\mu)}_{\Box_{ij}} = h^{(\mu)}_i h^{(\mu)}_j (h^{(\mu)}_i)^{-1}
(h^{(\mu)}_j)^{-1}
\end{equation}
is the holonomy of the gravitational connection around the square
loop $\Box_{ij}$,  considered over a face of the elementary cell,
each of whose sides has length $\mu V_o^{1/3}$ (where $\mu > 0$)
with respect to the flat fiducial metric $^o q_{ab}:=
\delta_{ij}\, ^o \omega^i_a\, ^o \omega^j_a $; fiducial triad $^o
e^a_k$ and cotriad $^o \omega^k_a$ satisfy $^o \omega^i_a\,^o
e^a_j = \delta^i_j$; the spatial part of the FRW metric is
$q_{ab}=a^2(t)\,^o q_{ab}$; $~Ar\,\Box_{ij}$ denotes the area of
the square; $V_o = \int_{\mathcal V} \sqrt{^o q} d^3 x$ is the
fiducial volume of $\mathcal V$. Figure 1 shows  geometrical setup
for determination of $h^{(\mu)}_{\Box_{ij}}$.

\begin{figure}[h]
\vspace{-0.02\textwidth} \hspace{-0.02\textwidth}
\includegraphics[width=0.55\textwidth]{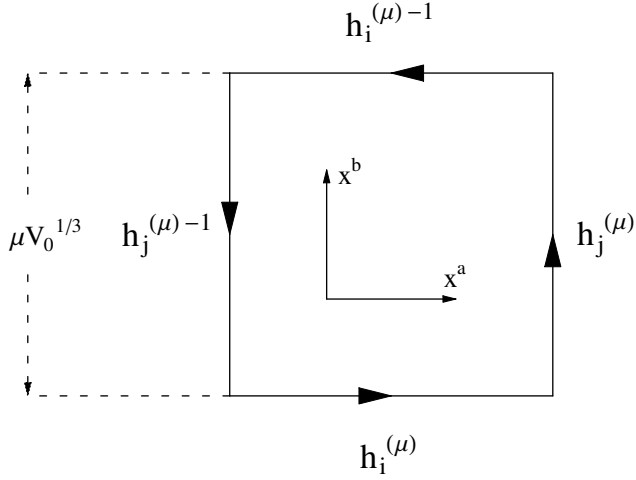}
\vspace{0.0\textwidth} \caption{Holonomy of connection around the
square loop. Suitable fiducial cotriads $^o \omega^k_a$ are chosen
to diagonalise  the connection $A^k_a$.}
\end{figure}

The holonomy along straight edge $ ^oe^a_k \partial_a $ of length
$\mu V_0^{1/3}$ reads
\begin{equation}\label{hol}
h^{(\mu)}_k (c) = \mathcal{P} \exp \,\big(\int_0^{\mu V_0^{1/3}}
\tau_{(k)} A^{(k)}_a dx^a \big) = \exp (\tau_{k}\mu c) =\cos (\mu
c/2)\;\id + 2\,\sin (\mu c/2)\;\tau_k,
\end{equation}
where $\tau_k = -i \sigma_k/2\;$ ($\sigma_k$ are the Pauli spin
matrices) and  $\mathcal{P}$ denotes the path ordering symbol.
Equation (5) presents the holonomy calculated in the fundamental,
j = 1/2, representation of SU(2). The {\it connection} $A^k_a$ and
the density weighted {\it triad} $E^a_k$ which occurs in
(\ref{identT}) is determined by the conjugate variables $c$ and
$p$ as follows: $A^k_a = \,^o\omega^k_a\,c\,V_0^{-1/3} \,$ and
$\,E^a_k = \,^oe^a_k\,\sqrt{q_o}\,p\,V_0^{-2/3} $, where $\,c =
\gamma \,\dot{a}\,V_0^{1/3}$ and $\,|p| = a^2\,V_0^{2/3}$.
Equation (\ref{hol}) presents the holonomy calculated  in the
fundamental, $j=1/2$, representation of $SU(2)$.

Making  use of (\ref{hamG}), (\ref{cur})  and the so-called
Thiemann identity \cite{TT}
\begin{equation}\label{identT}
\varepsilon_{ijk}\,e^{-1}\,E^{aj}E^{bk} = \frac{sgn(p)}{2\pi G
\gamma \mu V_0^{1/3}}\,\sum_k\,^o\varepsilon^{abc}\,
^o\omega^k_c\,Tr \Big(h_k^{(\mu)}\{(h_k^{(\mu)})^{-1},V\}\,\tau_i
\Big)
\end{equation}
leads to $H_g$ in the form
\begin{equation}\label{hamR}
    H_g = \lim_{\mu\rightarrow \,0}\; H^{(\mu)}_g ,
\end{equation}
where
\begin{equation}\label{hamL}
H^{(\mu)}_g = - \frac{sgn(p)}{2\pi G \gamma^3 \mu^3}
\sum_{ijk}\,N\, \varepsilon^{ijk}\, Tr \Big(h^{(\mu)}_i
h^{(\mu)}_j (h^{(\mu)}_i)^{-1} (h^{(\mu)}_j)^{-1}
h_k^{(\mu)}\{(h_k^{(\mu)})^{-1},V\}\Big),
\end{equation}
and where $V= |p|^{\frac{3}{2}}= a^3 V_0$ is the volume of the
elementary cell $\mathcal{V}$.

The classical total Hamiltonian for FRW universe with a massless
scalar field, $\phi$, reads
\begin{equation}\label{ham}
   H = H_g + H_\phi \approx 0,
\end{equation}
where $H_g$ is defined by (\ref{hamR}). The Hamiltonian of the
scalar field  is known to be: $H_\phi = N\,p^2_\phi
|p|^{-\frac{3}{2}}/2$, where $\phi$ and $p_\phi$ are the
elementary variables satisfying $\{\phi,p_\phi\} = 1$. The
relation $H \approx 0$ defines the {\it physical} phase space of
considered gravitational system with constraints.

Making use of (\ref{hol}) we calculate  (\ref{hamL}) and get the
{\it modified}  total Hamiltonian $H^{(\lambda)}_g$ corresponding
to (\ref{ham}) in the form
\begin{equation}\label{regH}
  H^{(\lambda)}/N = -\frac{3}{8\pi G \gamma^2}\;\frac{\sin^2(\lambda
\beta)}{\lambda^2}\;v + \frac{p_{\phi}^2}{2\, v},
\end{equation}
where
\begin{equation}\label{re1}
    \beta := \frac{c}{|p|^{1/2}},~~~v := |p|^{3/2}
\end{equation}
are the  canonical variables proposed in \cite{Ashtekar:2006wn}.
The variable $\beta = \gamma\dot{a}/a$ so it corresponds to the
Hubble parameter $\dot{a}/a$, whereas $v^{1/3} = a V_0^{1/3}$ is
proportional to the scale factor $a$. The relationship between the
coordinate length $\mu$  (which depends on $p$) and the physical
length $\lambda$ (which is a constant)  reads
\begin{equation}\label{kin}
    \lambda = \mu \,|p|^{1/2} = \mu\, a \,V_0^{1/3}.
\end{equation}

The complete Poisson bracket for the canonical variables
$(\beta,v,\phi,p_\phi)$ is defined to be
\begin{equation}\label{re2}
    \{\cdot,\cdot\}:= 4\pi G\gamma\;\bigg[ \frac{\partial \cdot}
    {\partial \beta} \frac{\partial \cdot}{\partial v} -
     \frac{\partial \cdot}{\partial v} \frac{\partial \cdot}{\partial \beta}\bigg] +
     \frac{\partial \cdot}{\partial \phi} \frac{\partial \cdot}{\partial p_\phi} -
     \frac{\partial \cdot}{\partial p_\phi} \frac{\partial \cdot}{\partial
     \phi}.
\end{equation}
The dynamics of a canonical variable $\xi$ is defined by
\begin{equation}\label{dyn}
    \dot{\xi} := \{\xi,H^{(\lambda)}\},~~~~~~\xi \in \{\beta,v,\phi,p_\phi\},
\end{equation}
where $\dot{\xi} := d\xi/d\tau$, and where $\tau$ is an evolution
parameter. The dynamics in the  {\it physical} phase space,
$\mathcal{F}_{phys}^{(\lambda)}$, is defined by solutions to
(\ref{dyn}) satisfying the condition $H^{(\lambda)}\approx 0$. The
solutions of (\ref{dyn}) ignoring the constraint
$H^{(\lambda)}\approx 0$ are in the  {\it kinematical} phase
space, $\mathcal{F}_{kin}^{(\lambda)}$.

In what follows we apply the Dirac method of dealing with
Hamiltonian constraints \cite{PAM}: the Poisson bracket is worked
out before one makes use of the constraint equations.

\section{Dynamics}

In  the case a Hamiltonian is a {\it constraint} which may be
rewritten in the form of a product of a simpler constraint and a
function on $\mathcal{F}_{kin}^{(\lambda)}$ which has no zeros,
the original dynamics may be reduced (to some extent) to the
dynamics with the simpler constraint.

Equation (\ref{regH}) can be rewritten as
\begin{equation}\label{product}
  H^{(\lambda)} = N\,H_0^{(\lambda)}\,\tilde{H}^{(\lambda)}\approx 0,
\end{equation}
where
\begin{equation}\label{defprod}
H_0^{(\lambda)} := \frac{3}{8 \pi G \gamma^2 v} \;\Big(\kappa
\gamma |p_\phi| + v\,\frac{|\sin(\lambda
\beta)|}{\lambda}\Big),~~~~~~ \tilde{H}^{(\lambda)}:= \kappa
\gamma |p_\phi| - v\, \frac{|\sin(\lambda \beta)|}{\lambda},
\end{equation}
where $\kappa^2 \equiv 4\pi G/3$.

\noindent It is clear that $H_0^{(\lambda)} = 0$ only in the case
when $p_\phi =0=\sin(\lambda \beta)$. Such case, due to
(\ref{1a})-(\ref{5a}), implies no dynamics.

\subsection{Relative dynamics}

An equation of motion for a function $f$ defined on {\it physical}
phase space,
 due to(\ref{dyn}), reads
\begin{equation}\label{funF}
    \dot{f} = \{f, N H_0^{(\lambda)}\tilde{H}^{(\lambda)}\} = \{f, N H_0^{(\lambda)}\}
    \tilde{H}^{(\lambda)} + N H_0^{(\lambda)}\{f,
    \tilde{H}^{(\lambda)}\} = N H_0^{(\lambda)}\{f,\tilde{H}^{(\lambda)}\},
    \end{equation}
since $\tilde{H}^{(\lambda)} = 0.$  By analogy, for other function
$g$ we have
\begin{equation}\label{funG}
    \dot{g} = \{g, N H_0^{(\lambda)}\tilde{H}^{(\lambda)}\} =
    N H_0^{(\lambda)}\{g,\tilde{H}^{(\lambda)}\},~~~~~
    \mbox{for}~~~~~\tilde{H}^{(\lambda)}
    \approx 0.
\end{equation}
Therefore we have the relation
\begin{equation}\label{fraC}
    \frac{\dot{f}}{\dot{g}} = \frac{df}{dg} = \frac{N H_0^{(\lambda)}
    \{f,\tilde{H}^{(\lambda)}\}}
    {N H_0^{(\lambda)}\{g,\tilde{H}^{(\lambda)}\}} = \frac{\{f,\tilde{H}^{(\lambda)}\}}
    {\{g,\tilde{H}^{(\lambda)}\}}
    ,~~~~\mbox{as}~~~~H_0^{(\lambda)} \neq 0,
\end{equation}
which we rewrite in the form
\begin{equation}\label{integ}
    \frac{df}{\{f,\tilde{H}^{(\lambda)}\}} = \frac{dg}{\{g,\tilde{H}^{(\lambda)}\}}
\end{equation}
Equation (\ref{integ}) shows that in the case of the {\it
relative} dynamics we have: (i) one phase space variable may be
used as an `evolution parameter' of all other variables, (ii)
dynamics is gauge independent in the sense that there is no
dependance on the specific choice of the lapse function $N$, and
(iii) suitable choice of $N$ may lead to a simpler form of
Hamiltonian.

\subsection{Solution of the relative dynamics}

Since the relative dynamics is gauge independent, it is reasonable
to chose the gauge which simplifies the calculations. Our choice
is $N := 1/H_0^{(\lambda)}$. In this gauge the equations of motion
read
\begin{eqnarray}\label{1a}
  \dot{p_{\phi}}&=&0, \\\ \label{2a}
  \dot{\beta}&=&  -4\pi G\gamma
  \;\frac{|\sin(\lambda\, \beta)|}{\lambda},
  \\ \label{3a}
  \dot{\phi}&=& \kappa\gamma~\textrm{sgn}(p_{\phi}), \\ \label{4a}
  \dot{v}&=& 4\pi G\gamma v \cos(\lambda\,
  \beta)~\textrm{sgn}(\sin(\lambda\, \beta)), \\ \label{5a}
  \tilde{H}^{(\lambda)} & = & 0.
\end{eqnarray}
Combining (\ref{3a}) with (\ref{4a})  gives
\begin{eqnarray}\label{pp1}
\frac{\dot v}{\dot \phi} =  3\kappa v \cos{(\lambda
\beta)}\;\textrm{sgn}(\sin{(\lambda
\beta)})\;\textrm{sgn}(p_{\phi}).
\end{eqnarray}
Rewriting (\ref{pp1}) (and using $\dot v/\dot \phi = dv/d\phi$)
gives
\begin{eqnarray}\label{zcos}
\frac{\textrm{sgn}(\sin(\lambda \beta))}{\cos(\lambda
\beta)}\;\frac{dv}{v} = 3\kappa\;\textrm{sgn}(p_{\phi})\;d\phi
\end{eqnarray}
Making use of the identity $\;\sin^2(\lambda \beta)+\cos^2(\lambda
\beta)=1$ and (\ref{5a}) gives
\begin{eqnarray}\label{nacos}
|\cos{(\lambda \beta)}|= \sqrt{1-\Big(\frac{\kappa \gamma p_{\phi}
\lambda }{v}}\Big)^2
\end{eqnarray}
Combining (\ref{zcos}) with (\ref{nacos}), for $\beta\in ]0, \pi/2
\lambda[$, leads to
\begin{equation}\label{dynamika}
    \frac{dv}{\sqrt{
   v^2-(\kappa\gamma\lambda
   p_\phi)^2}}= 3 \kappa\; \textrm{sgn}(p_{\phi})\;d\phi .
\end{equation}
Since $p_\phi$ is just a constant (due to (\ref{1a})) we can
easily integrate (\ref{dynamika}) and get
\begin{equation}\label{int}
    \ln\bigg|v+\sqrt{v^2-(\kappa\gamma\lambda
   p_\phi)^2}\bigg|=3 \kappa\;\textrm{sgn}(p_{\phi})(\phi-\phi_0) .
\end{equation}
Rewriting (\ref{int}) leads to
\begin{equation}\label{res2}
2\,v = \exp{\big(3\kappa\;\textrm{sgn}(p_{\phi})\;(\phi -
\phi_{0})\big)} + (\kappa\gamma |p_{\phi}|
\lambda)^2\cdot\exp{\big(-3\kappa\;\textrm{sgn}(p_{\phi})\;(\phi -
\phi_{0})\big)}.
\end{equation}
The solution for the variable $\beta$ may be easily determined
from (\ref{5a}) rewritten as
\begin{equation}\label{res3}
\kappa \gamma |p_\phi| = v\, \frac{|\sin(\lambda \beta)|}{\lambda}
\end{equation}
The final expression reads
\begin{equation}\label{res4}
    \sin(\lambda \beta)= \frac{2 \kappa\gamma\lambda
   |p_\phi|}{\exp\big(3\kappa\, \textrm{sgn}(p_{\phi})\,(\phi-\phi_0)\big)+
   (\kappa\gamma\lambda
   p_\phi)^2\exp \big(-3\kappa\, \textrm{sgn}(p_{\phi})\,(\phi-\phi_0)\big)}
\end{equation}
where the domain of the variable $\beta$ has been extended to the
interval $ ]0,\pi/\lambda[$.

Equations (\ref{res2}) and (\ref{res4}) present the dependence of
the canonical variables $v$ and $\beta$ on the evolution parameter
$\phi$, which is a monotonic function due to (\ref{3a}).

\section{Algebra of observables}

A function, $\mathcal{O}$, defined on phase space   is a Dirac
observable  if
\begin{equation}\label{dirac}
\{\mathcal{O},H^{(\lambda)}\} \approx 0.
\end{equation}
Since we have
\begin{equation}\label{Dirac}
\{\mathcal{O},H^{(\lambda)}\}=  \{\mathcal{O},N H_0^{(\lambda)}
\tilde{H}^{(\lambda)}\}= N H_0^{(\lambda)}\{\mathcal{O},
\tilde{H}^{(\lambda)}\} + \{\mathcal{O}, N
H_0^{(\lambda)}\}\tilde{H}^{(\lambda)},
\end{equation}
it is clear that on the constraint surface,
$\tilde{H}^{(\lambda)}=0$, the Dirac observable satisfies
(independently on the choice of $N$) a much simpler equation
\begin{equation}\label{donot}
\{\mathcal{O},\tilde{H}^{(\lambda)}\}\approx 0.
\end{equation}
Thus,  we put $N:= 1/H_0^{(\lambda)}$ and solve (\ref{dirac}) in
the whole phase space, i.e. we solve the equation
\begin{equation}\label{dir}
\frac{\sin(\lambda\beta)}{\lambda}\,\frac{\partial
\mathcal{O}}{\partial\beta} - v \cos(\lambda\beta)\,\frac{\partial
\mathcal{O}}{\partial v} - \frac{\kappa\,\textrm{sgn}(p_{\phi})}{4
\pi G}\,\frac{\partial \mathcal{O}}{\partial\phi} = 0.
\end{equation}

A function $\mathcal{O} =
\mathcal{O}(\mathcal{O}_1,\ldots\mathcal{O}_k)$ satisfies
(\ref{dir}) if
\begin{equation}\label{obser}
\{\mathcal{O}_1,\tilde{H}^{(\lambda)}\} = 0=
\{\mathcal{O}_2,\tilde{H}^{(\lambda)}\} =\ldots =
\{\mathcal{O}_k,\tilde{H}^{(\lambda)}\},
\end{equation}
where $k+1$ is the dimension  of the {\it kinematical} phase
space. It is so because one has
\begin{equation}\label{prop}
\{\mathcal{O},\tilde{H}^{(\lambda)}\} = \frac{\partial
\mathcal{O}}{\partial
\mathcal{O}_1}\,\{\mathcal{O}_1,\tilde{H}^{(\lambda)}\}  + \ldots
+ \frac{\partial \mathcal{O}}{\partial
\mathcal{O}_k}\,\{\mathcal{O}_k,\tilde{H}^{(\lambda)}\}.
\end{equation}

In what follows we consider only {\it elementary} observables. The
set of such observables, $\mathcal{E}$, is defined by the
requirements: (i) each element of $\mathcal{E}$ is  a solution to
(\ref{dir}), (ii) elements of $\mathcal{E}$ are functionally
independent on the constraint surface, $\tilde{H}^{(\lambda)}=0$,
(iii) elements of $\mathcal{E}$ satisfy a Lie algebra, and (iv)
two sets of observables satisfying two algebras are considered to
be the same if these algebras are isomorphic.

In our case $k = 3$ and  solutions to (\ref{dir}) are found to be
\begin{equation}\label{obser1}
\mathcal{O}_1:= p_{\phi},~~~\mathcal{O}_2:= \phi -
\frac{s}{3\kappa}\;\textrm{arth}\big(\cos(\lambda \beta)\big),~~~~
\mathcal{O}_3:= s\,v\, \frac{\sin(\lambda \beta)}{\lambda},
\end{equation}
where $s := \textrm{sgn}(p_\phi)$. One may verify that the
observables satisfy the Lie algebra
\begin{equation}\label{ala1}
\{\mathcal{O}_2,\mathcal{O}_1\}=
1,~~~~\{\mathcal{O}_1,\mathcal{O}_3\}= 0,~~~~
\{\mathcal{O}_2,\mathcal{O}_3\}=  \gamma\kappa .
\end{equation}

Because of the constraint $\tilde{H}^{(\lambda)}=0$ (see
(\ref{res3})), we have
\begin{equation}\label{con}
\mathcal{O}_3=  \gamma \kappa \,\mathcal{O}_1.
\end{equation}
Thus,  we have only two elementary Dirac observables which may be
used to parameterize  the physical phase space
$\mathcal{F}_{phys}^{(\lambda)}$. To identify the Poisson bracket
in $\mathcal{F}_{phys}^{(\lambda)}$ consistent with the Poisson
bracket (\ref{re2}) defined in $\mathcal{F}_{kin}^{(\lambda)}$, we
find a symplectic twoform corresponding to (\ref{re2}). It reads
\begin{equation}\label{sym1}
    \omega = \frac{1}{4 \pi G \gamma} d\beta \wedge d v + d\phi
    \wedge dp_\phi .
\end{equation}
The twoform $\omega$ is degenerate on
$\mathcal{F}_{phys}^{(\lambda)}$ due to the constraint
$\tilde{H}^{(\lambda)}=0$. Making use of the explicit form of this
constraint (\ref{res3}) and  the functional form of
$\mathcal{O}_1$ and $\mathcal{O}_2$, leads to the symplectic form
$\Omega$ on $\mathcal{F}_{phys}^{(\lambda)}$. Direct calculations
give (see App. B)
\begin{equation}\label{sym2}
    \Omega := \omega_{| \tilde{H}^{(\lambda)}=0} = d\, \mathcal{O}_2
    \wedge d\, \mathcal{O}_1 ,
\end{equation}
where $\omega_{| \tilde{H}^{(\lambda)}=0}$ denotes the reduction
of $\omega$ to the constraint surface. The Poisson bracket
corresponding to (\ref{sym2}) reads
\begin{equation}\label{sym3}
\{\cdot,\cdot\}:=\frac{\partial\cdot}{\partial
\mathcal{O}_2}\frac{\partial\cdot}{\partial \mathcal{O}_1} -
\frac{\partial\cdot}{\partial
\mathcal{O}_1}\frac{\partial\cdot}{\partial \mathcal{O}_2}
\end{equation}
so the algebra satisfied by $\mathcal{O}_1$ and $\mathcal{O}_2$
has a simple form given by
\begin{equation}\label{sym4}
\{\mathcal{O}_2,\mathcal{O}_1\}= 1.
\end{equation}

Our kinematical phase space, $\mathcal{F}_{kin}^{(\lambda)}$, is
four dimensional. In {\it relative} dynamics one variable is used
to parametrize three others. Since the constraint relates  the
variables, we have only two independent variables. This is the
reason  we have only two elementary physical observables
parametrizing $\mathcal{F}_{phys}^{(\lambda)}$.

\section{Functions on  phase space}

In this section we discuss the functions on the {\it constraint}
surface that may describe singularity aspects of our cosmological
model. Considered functions are not observables, but they can be
expressed in terms of observables and an evolution parameter
$\phi$. They do become observables for each fixed value of $\phi$,
since in such case they are only functions of observables.

\subsection{Energy density}

An expression for the energy density $\rho$ of the scalar field
$\phi$ reads
\begin{equation}\label{rho2}
\rho(\lambda,\phi)=\frac{1}{2}\,\frac{p_{\phi}^2}{v^2}.
\end{equation}
In terms of elementary observables we have
\begin{equation}\label{obser}
p_\phi = \mathcal{O}_1,~~~~v = \kappa\gamma\lambda\,
    |\mathcal{O}_1|\,\cosh\big(3\kappa  (\phi- \mathcal{O}_2)\big).
\end{equation}
For fixed $p_\phi$ the density $\rho$ takes its maximum value at
the minimum value of $v$. Rewriting (\ref{res2}) in the form
\begin{equation}\label{id1}
\frac{v}{\triangle}= \cosh\big(3 \kappa s (\phi - \phi_{0}) - \ln
\triangle\big),~~~~~\mbox{where}~~~~~\triangle:=
\kappa\gamma\lambda\,|p_\phi|,
\end{equation}
we can see that  $\,\cosh(\cdot)\,$  takes minimum value equal to
one at $ 3 \kappa s\,(\phi - \phi_{0})= \ln \triangle$. Thus, the
maximum value of the density, $\rho_{\max}$, corresponds to $v =
\triangle$ and  reads
\begin{equation}\label{cr1}
\rho_{\max} = \frac{1}{2\kappa^2 \gamma^2}\,\frac{1}{ \lambda^2}.
\end{equation}

We can determine $\rho_{\max}$  if we know $\lambda$. However,
$\lambda$ is a free parameter of the formalism. Thus, finding the
critical energy density of matter corresponding to the big bounce
is an open problem.

It is tempting to apply (\ref{cr1}) to the Planck scale. To make
use of Planck's length $l_{Pl}:= \sqrt{\hbar G/c^3}$ and Planck's
energy density $\rho_{Pl}:= c^5/\hbar G^2$, we multiply
(\ref{cr1}) by $c^2$ and recall that $\kappa^2 \equiv 4\pi G/3$.
Thus (\ref{cr1}) reads
\begin{equation}\label{ps1}
 \rho_{\max} = \frac{3\; c^2}{8 \pi G
 \gamma^2}\;\frac{1}{\lambda^2}.
\end{equation}
Substituting $\lambda = l_{Pl}$ into (\ref{ps1}) gives $\rho_{max}
=3/8 \pi \gamma^2\;\rho_{Pl}\simeq 2,07\;\rho_{Pl}$. Resolving
(\ref{ps1}) in terms of $\lambda$ makes possible finding $\lambda$
corresponding to $\rho_{Pl}$. We get $\lambda= \sqrt{3/8 \pi
\gamma^2}\;l_{Pl}\simeq 1,44\;l_{Pl}$. (We have used $\gamma\simeq
0.24$ determined in black hole entropy calculations
\cite{Domagala:2004jt,Meissner:2004ju}.) Surprisingly, the
classical expression (\ref{ps1}) fits the Planck scale.

A natural next step is an examination of the energy scale of the
big bounce at the quantum level. Preliminary calculations,
\cite{Malkiewicz:2009zd}, suggest that the energy scale would be
described by the classical expression (\ref{ps1}).

\subsection{Geometrical operators}

In the case the volume $V=a^3 V_0$ of an elementary cell
$\mathcal{V}$ is a cube,  geometrical operators have simple
dependence on canonical variables. Since the volume operator, $V$,
is given by
\begin{equation}\label{g3}
    V(\phi) =  v,
\end{equation}
the area operator, $A$, reads
\begin{equation}\label{g2}
    A(\phi) = v^{2/3},
\end{equation}
and the  length operator, $L$, is found to be
\begin{equation}\label{g1}
    L(\phi) = v^{1/3}.
\end{equation}

As far as we know, the geometrical operators have been considered
so far only in the {\it kinematical} Hilbert space of the loop
quantum {\it gravity}
\cite{Thiemann:1996at,Bianchi:2008es,Rovelli:1994ge,Ashtekar:1996eg,Ashtekar:1997fb}.
We propose the examination of these operators in the {\it
physical} Hilbert space of the loop quantum {\it cosmology}.

It results from (\ref{res3}) that we have
\begin{equation}\label{mins}
    v_{\min} = \kappa\gamma\lambda\,|p_\phi|
\end{equation}
so  the geometrical operators are bounded from below by zero (as
$\lambda\,|p_\phi| \rightarrow 0$).

An examination of the spectra of the geometrical operators at the
quantum level is the next natural step.  An interesting question
is: Do these operators have the  nonzero minimum and discrete
eigenvalues?  It is expected that the story will turn out to be
similar to the case of the linear harmonic oscillator, where we
have the nonzero ground-state energy and discrete energy levels.
We present an answer to this intriguing question in our
forthcoming paper \cite{all}.

\section{Conclusions}

We have shown that the resolution of the initial singularity of
the flat FRW model with massless scalar field is due to the
modification of the model at the {\it classical} level by making
use of the loop geometry. The modification is parametrized by the
{\it continuous} parameter  $\lambda $. Each value of $\lambda$
specifies the critical energy density of the scalar field
corresponding to the big bounce.  As there is no specific choice
of $\lambda$, the BB may occur at any low and high densities. The
former case (big $\lambda$) contradicts the data of observational
cosmology (there was no BB in the near past!) and leads to weakly
controlled modification  of the expression for the curvature
$F^k_{ab}$, i.e. gravitational part of the Hamiltonian (see the
appendix). The latter case (small $\lambda$) gives much better
approximation for the classical Hamiltonian (see the appendix),
but may easily lead to densities much higher than the Planck scale
density, where the classical formalism is believed to be
inadequate. Finding specific value of the parameter $\lambda$,
i.e. the {\it energy scale} specific to BB is an open problem.

Our approach is quite different from the so-called effective or
polymerization method (see, e.g. \cite{Mielczarek:2008zv}), where
the replacement $\beta \rightarrow \sin(\lambda \beta)/\lambda$ in
the Hamiltonian finishes the procedure of quantization. In our
method this replacement has been done entirely at the {\it
classical} level: Eq. (10) results from using an explicit form of
the holonomy (5) in (8). Quantization consists in finding a
self-adjoint representation of observables on the physical phase
space and an examination of the spectra of these observables
\cite{Malkiewicz:2009zd,Malkiewicz:2009xz}.

The elementary observables $\mathcal{O}_1$ and $\mathcal{O}_2$
constitute a complete set of constants of motion in the constraint
surface. They are used to parametrize the physical phase space and
are ``building blocks'' for the compound observables like the
energy density of the scalar field and the geometrical operators.
So they have deep physical meaning. The role of $\mathcal{O}_1$
and $\mathcal{O}_2$ becomes even more important at the quantum
level as they enable finding quantum operators corresponding to
the classical compound observables \cite{Malkiewicz:2009qv}.

Our theoretical framework may be used for examination of the {\it
discreteness} aspects of geometrical {\it quantum} operators,
which may help in the selection of $\lambda$. An extension of our
formalism to the quantum level is straightforward. The algebra of
observables is defined in the physical phase space (hyper-surface
in the kinematical phase space determined by the constraint
equation). The carrier space of the self-adjoint representation of
the algebra defines the {\it physical} Hilbert space. Examination
of the eigenvalue problem of the length, area and volume operators
(for fixed value of an evolution parameter they are observables)
may lead to the specification of an unique $\lambda$ (or an
interval of allowed values). Our next paper \cite{all} is devoted
to examination of these problems.

It may happen, however,  that the value of the parameter $\lambda$
cannot be  determined, for some reason, theoretically. The story
may turn out to be similar to the case of the short-range {\it
repulsive} part of the potential of the nucleon-nucleon
interaction introduced to explain the scattering data \cite{RJ}
and the nuclear matter saturation of energy \cite{BM}. In such a
case $\lambda$ will become a phenomenological variable
parameterizing our ignorance of microscopic properties  of the
universe.

An independent source of information on discreteness aspects of
geometry is the observational cosmology. The cosmic projects for
the detection of gamma ray bursts   may reveal that the velocity
of cosmic photons depend on their wave lengths, which may be
ascribed to the {\it foamy} nature of  spacetime
\cite{RB,Albert:2007qk,Lamon:2008es}. The detection of the
primordial gravitational waves created at  the big bounce may
bring valuable information on the geometry  of this phase
\cite{Mielczarek:2008pf,Calcagni:2008ig,Grain:2009kw,Grain:2009cj}.
The observational cosmology data may help to determine the
phenomenological value of the parameter $\lambda$.

\begin{acknowledgments}
We are grateful to Janusz D\c{a}browski, Jacek Jezierski, and
Andrzej Kr\'{o}lak for helpful discussions. We would like to thank
the anonymous referee for the suggestion of including
\cite{Bojowald:2001xe,Date:2004fj} into our list of references.
\end{acknowledgments}

\appendix

\section{Holonomy corrections}

The curvature  of  $SU(2)$ connection $F^k_{ab} =
\partial_a A^k_b - \partial_b A^k_a + \epsilon^k_{ij} A^i_a
A^j_b$, entering the expression (\ref{hamG}) for the gravitational
part of the Hamiltonian, can be expressed in terms of holonomies.
Using the mean-value and Stokes' theorems we
have\begin{equation}\label{stokes}
    \tau_k\,F^k_{ab}(\vec{x})\approx\frac{1}{s^\sigma_{ab}}\int_\sigma\,
    \tau_k\,F^k_{cd}\,d x^c\wedge
    dx^d\approx\frac{1}{s^\sigma_{ab}}\Big( \mathcal{P}\exp\big(\oint_{\partial \sigma}
    \tau_k\,A^k_c\,dx^c\big) - 1\Big),
\end{equation}
where $\partial \sigma$ is the boundary of a small surface
$\sigma$ with center at $\vec{x}$, and where $s^\sigma_{ab}:=
\int_\sigma dx^a\wedge dx^b$. The expression for $F^k_{ab}$ is
exact  but in the limit when we shrink the area enclosed by the
loop $\partial \sigma$ to zero. If we choose $\partial \sigma$ in
the form of the square  $\square_{ij}$ with sides length $\mu$,
the expression for a {\it small} value of $\mu = \mu_0$ has the
form \cite{Mielczarek:2008zz}
\begin{equation}\label{finite}
F^k_{ab}(\mu_0)= \lim_{\mu\,\rightarrow \,\mu_0}
\Big\{-2\;Tr\;\Big(\frac{h^{(\mu)}_{\Box_{ij}}-1}{\mu^2
V_o^{2/3}}\Big)\;{\tau^k}\; ^o\omega^i_a  \; ^o\omega^j_a +
\frac{\mathcal{O}(\mu^4)}{\mu^2}\Big\},
\end{equation}
and we have
\begin{equation}\label{zero}
F^k_{ab}= \lim_{\mu_0\,\rightarrow \,0}\, F^k_{ab}(\mu_0).
\end{equation}

In the standard LQC  the $\mathcal{O}(\mu^4)$ holonomy corrections
are ignored (see, e.g. \cite{Ashtekar:2006wn,Ashtekar:2007em}). It
was found in \cite{Mielczarek:2008zz,Hrycyna:2008yu} that
including higher order corrections  leads to new curvature
singularities different from the initial singularity and increases
an ambiguity problem of loop  cosmology. However, the holonomy
corrections do not change  the result that the big bounce is a
consequence of the loopy nature of geometry \cite{Chiou:2009hk}.

Taking only the first term of (\ref{finite}) leads to the simplest
modification of gravity, but may be insufficient for the
description of the inflationary phase. The choice of $\mu_0$ based
on the expectation that Big Bounce should occur at the Planck
scale \cite{Ashtekar:2006wn} has little justification
\cite{Bojowald:2008ik}. The significance of Planck's scale for
quantum gravity seems to be rather a belief than proved result
(see, e.g. \cite{Meschini:2006gm}). Heuristic reasoning playing
game at the same time with Heisenberg's uncertainty principle,
Schwarzschild's radius and process of measurement cannot replace a
proof  (see, e.g. \cite{WP}).

\section{Symplectic form}

The symplectic form on the physical phase space $\Omega$ may be
obtained from the symplectic form on the kinematical phase space
$\omega$ by taking into account the constraint (\ref{res3}).

The symplectic form corresponding to (\ref{re2}) reads
\begin{eqnarray}\label{form1}
\omega = d\phi\wedge dp_{\phi} + \frac{1}{4\pi
G\gamma}\,d\beta\wedge dv= d\phi\wedge dp_{\phi} +
\frac{1}{3\kappa^2\gamma}\,d\beta\wedge dv .
\end{eqnarray}
Making use of (\ref{res3}) we get
\begin{equation}\label{ff2}
dv=
\frac{\kappa\gamma\lambda\,\,\textrm{sgn}(p_{\phi})}{\sin(\lambda\beta)}\,dp_{\phi}
- \lambda\,\textrm{ctg}(\lambda\beta)\,d\beta .
\end{equation}
Insertion of (\ref{ff2}) into (\ref{form1}) gives
\begin{eqnarray}
\Omega = d\phi\wedge dp_{\phi} +
\frac{\textrm{sgn}(p_{\phi})}{3\kappa}\frac{\lambda}{\sin(\lambda\beta)}\,d\beta\wedge
dp_{\phi}.
\end{eqnarray}
Since
\begin{eqnarray}
\frac{\lambda}{\sin(\lambda\beta)}= -
\frac{d\,\textrm{arth}(\cos{\lambda\beta})}{d\beta} ,
\end{eqnarray}
we have
\begin{eqnarray}
\Omega = \bigg(d\phi - \frac{\textrm{sgn}(p_{\phi})}{3\kappa}\,
\frac{d\,\textrm{arth}(\cos{\lambda\beta})}{d\beta}\,d\beta
\bigg)\wedge dp_{\phi} .
\end{eqnarray}
On the other hand, due to (\ref{obser1}), we have
\begin{eqnarray}
\mathcal{O}_1= p_{\phi},\,\,\,\,\, \mathcal{O}_2= \phi -
\frac{\textrm{sgn}(p_{\phi})}{3\kappa}\,\textrm{arth}(\cos(\lambda\beta))
.
\end{eqnarray}
Therefore,
\begin{eqnarray}
\Omega = d\mathcal{O}_{2}\wedge d\mathcal{O}_{1}.
\end{eqnarray}
Thus, the physical phase space may be parametrized by the
variables $\mathcal{O}_{1}$ and $\mathcal{O}_{2}$, and the
corresponding Poisson bracket is given by (\ref{sym3}).

\end{document}